\documentstyle[twoside,fleqn,npb,epsfig]{article}

\begin{document}

%
% put your own definitions here:
%   \newcommand{\cZ}{\cal{Z}}
%   \newtheorem{def}{Definition}[section]
%   ...
\def\beq{\begin{equation}} 
\def\eeq{\end{equation}} 
\def\qb{{\bar q}}
\def\MB{{\cal M}_B}
\def\beqn{\begin{eqnarray}} 
\def\eeqn{\end{eqnarray}} 

\def\lq{\left[} 
\def\rq{\right]} 
\def\rg{\right\}} 
\def\lg{\left\{} 
\def\({\left(} 
\def\){\right)} 

\newcommand{\ttbs}{\char'134}
\newcommand{\AmS}{{\protect\the\textfont2
  A\kern-.1667em\lower.5ex\hbox{M}\kern-.125emS}}

% add words to TeX's hyphenation exception list
\hyphenation{author another created financial paper re-commend-ed}

% declarations for front matter

\title{QCD corrections to vector boson fusion processes\thanks{Talk 
given by Dieter Zeppenfeld}}

\author{T. Figy,\address{Department of Physics, University of Wisconsin, 
        Madison, WI 53706, U.S.A.}
%        \thanks{Footnotes should appear on the first page only to
%                indicate your present address (if different from your
%                normal address), research grant, sponsoring agency, etc.
%                These are obtained with the {\tt\ttbs thanks} command.}
        C. Oleari\address{Dipartimento di Fisica "G. Occhialini", 
        Universit\`a di Milano-Bicocca, %Piazza della Scienza, 3,
        20126 Milano, Italy}
        and
        D. Zeppenfeld\address{Institut f\"ur Theoretische Physik, 
        Universit\"at Karlsruhe, Postfach, 76128 Karlsruhe, Germany}
%        X.-Y. Wang\address{Economics Department, University of Winchester, \\
%        2 Finch Road, Winchester, Hampshire P3L T19, United Kingdom}
}

\begin{abstract}
NLO QCD corrections to $H$, $W$ and $Z$ production via vector boson fusion
have recently been calculated in the form of flexible parton level 
Monte Carlo programs. This allows for the calculation of distributions
and cross sections with cuts at NLO accuracy. Some features of the 
calculation, as well as results for the LHC, are reviewed.
\end{abstract}

\maketitle

\section{Introduction}
\begin{figure}[bht] 
\vspace*{-0.5in}
\centerline{ 
\epsfig{figure=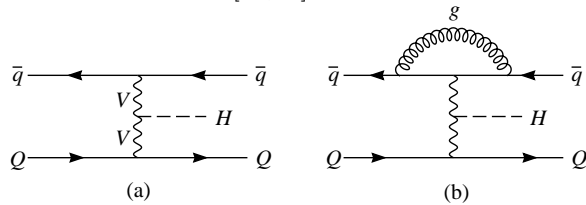,width=0.5\textwidth,clip=} \ \  
} 
\vspace*{-0.3in}
\caption{ 
\label{fig:feyn2} 
Feynman graphs contributing to the VBF process $\bar qQ\to \bar qQH$ at
(a) tree level and (b) including virtual corrections to the upper quark line.
}
\vspace*{-0.3in}
\end{figure}

Vector-boson fusion (VBF) processes have emerged as a particularly
interesting class of scattering events from which one hopes to gain
insight into the dynamics of electroweak symmetry breaking. The most
prominent example is Higgs boson production via VBF, which is shown in
Fig.~\ref{fig:feyn2}. This process has been studied intensively as a
tool for Higgs boson discovery~\cite{ATLAS-CMS,wbfhtautau,wbfhtoww} and the
measurement of Higgs boson couplings~\cite{Zeppenfeld:2000td} in
$pp$ collisions at the CERN Large Hadron Collider (LHC). The two scattered
quarks in a VBF process are usually visible as forward jets and greatly help
to distinguish these $Hjj$ events from backgrounds.

Analogous to Higgs boson production via VBF, the production of $Wjj$ and 
$Zjj$ events via vector-boson fusion will proceed with sizable cross section 
at the LHC. These 
processes have been considered previously at leading order for the study
of rapidity gaps at hadron 
colliders~\cite{Chehime:1992ub,Rainwater:1996ud,Khoze:2002fa}, 
as a background to Higgs boson searches in VBF~\cite{wbfhtautau,wbfhtoww}, 
or as a probe of anomalous triple-gauge-boson couplings~\cite{Baur:1993fv}, 
to name but a few examples. 

In order to match the achievable statistical precision in such LHC studies,
the inclusion of QCD corrections in the predicted VBF cross sections is 
required. While NLO QCD corrections to the Higgs boson total cross section
have  
been known for over a decade~\cite{WBF_NLO}, NLO corrections to distributions
have become available only recently, both for Higgs boson
production~\cite{Figy:2003nv,Berger:2004pc} and for $l^+l^-$ and $l\nu$ 
production in VBF~\cite{Oleari:2003tc}. These new calculations
use the subtraction method of Catani and Seymour~\cite{CS} 
to construct flexible 
parton level Monte Carlo programs for the calculation of NLO corrected 
distributions. In this talk we describe the work of 
Refs.~\cite{Figy:2003nv,Oleari:2003tc}.

%\vspace*{-0.3in}

\section{Elements of the calculation}
Because of the color singlet exchange in the LO VBF diagrams, any 
${\cal O}(\alpha_s)$ corrections, where gluons are simultaneously 
attached to both the upper and the lower quark line in Fig.~1,
vanish identically. Hence it is sufficient to consider radiative 
corrections to a single quark line at a time. For Higgs boson production,
the virtual diagrams reduce to simple vertex corrections, like the one 
depicted in Fig.~1(b).
For $W$ and $Z$ production also box graphs appear in
corrections to diagrams where the final state vector boson is
emitted from one of the quark lines.

The soft and collinear singularity structure for real emission
corrections only depends on the color of the external partons and,
therefore, is universal for
all VBF processes. Consider, for example, the emission of a gluon from
the upper quark line in Fig.~\ref{fig:feyn2}(a), with momenta given by
\beq
\bar q(p_a) + Q(p_b) \to g(p_1) + \bar q(p_2) + Q(p_3) + B(P)\; .
\eeq
Here $B=H,W,Z$ denotes the produced boson and we write the tree-level
amplitude 
for the emission process as
${\cal M}^{\bar q}_r={\cal M}^{\bar q}_r(p_a,p_1,p_2;q)$, where 
$q=p_1+p_2-p_a$ is the four momentum of 
the vector boson $V$ which is attached to the upper quark line
and which has virtuality $Q^2=-q^2$.
The singularities of the 3-parton phase-space integral of 
$|{\cal M}^{\bar q}_r|^2$ can be absorbed into a single counter term
\beq        % checked 5/14
\label{eq:Mqsing}
\left|{\cal M}^{\bar q}\right|^2_{\rm sing} 
%=  {\cal D}_2^{{\bar q} 1}+{\cal D}_{12}^{\bar q} 
% = 8\pi\alpha_s(\mu_R)\, \frac{4}{3}\, \frac{1}{Q^2}\,
%     \frac{x^2+z^2}{(1-x)(1-z)} \left|\MB^\qb\right|^2\,,
 = \frac{32\pi\alpha_s(\mu_R)}{3\; Q^2}\,
     \frac{x^2+z^2}{(1-x)(1-z)} \left|\MB^\qb\right|^2,
\eeq
where $\MB^\qb=\MB^\qb(\tilde p_a,\tilde p_2;q)$ is the Born amplitude 
for the lowest order process 
\beq
\bar q(\tilde p_a) + Q(p_b) \to \bar q(\tilde p_2) + Q(p_3) + B(P)\;,
\eeq
evaluated at the phase-space point 
\beq        % checked 5/14
\tilde p_a = x p_a\;,\qquad \tilde p_2 = p_1+p_2-(1-x)p_a\;,
\eeq
with
\beqn        % checked 5/14
\label{eq:defxz1}
      x &=& 1- \frac{p_1\cdot p_2}{(p_1+p_2)\cdot p_a} \,, \\
      z &=& 1-\frac{p_1\cdot p_a}{(p_1+p_2) \cdot p_a} =
              \frac{p_2\cdot p_a}{(p_1+p_2) \cdot p_a}\,.
\label{eq:defxz}
\eeqn
This choice continuously interpolates between the singularities due to
final-state soft gluons ($p_1\to 0$ corresponding to $x\to 1$ and $z\to 1$), 
collinear final-state partons ($p_1||p_2$ resulting in $p_1\cdot p_2\to 0$ 
or $x\to 1$) and gluon emission collinear to the initial-state anti-quark 
($p_1\to (1-x)p_a$ and $z\to 1$). The subtracted real emission amplitude
squared, $|{\cal M}^\qb_r|^2-|{\cal M}^\qb|^2_{\rm sing}$, leads to a 
finite phase-space integral of the real parton emission cross section,
and these integrals are evaluated numerically in $D=4$ dimensions.

The singular counter terms are integrated analytically, in $D=4-2\epsilon$
dimensions, over the phase space of the collinear and/or soft final-state 
parton. $1/\epsilon$ terms proportional to the $P^{qq}$ splitting
function disappear after factorization of the parton distribution
functions. The remaining divergent terms in the integral of 
Eq.~(\ref{eq:Mqsing}) yield the contribution 
(we are using the notation of Ref.~\cite{CS} and $C_F=4/3$)  
\beqn
\label{eq:I}
<{\boldmath I}(\epsilon)> &=& |\MB^{\bar q}|^2 \frac{\alpha_s(\mu_R)}{2\pi} C_F
\(\frac{4\pi\mu_R^2}{Q^2}\)^\epsilon \nonumber \\
&\times& \Gamma(1+\epsilon)
\lq\frac{2}{\epsilon^2}+\frac{3}{\epsilon}+9-\frac{4}{3}\pi^2\rq .
\eeqn

The  $1/\epsilon^2$ and $1/\epsilon$ soft and collinear
divergences cancel against the poles of the virtual 
corrections. For Higgs boson production they are depicted in
Fig.~\ref{fig:feyn2}(b) and are given by a simple vertex correction
only, which is proportional to the Born amplitude. For the more general
case of $W$ and $Z$ production, the virtual
amplitudes involve box contributions and have a much more complex
structure. However, by isolating the $1/\epsilon^2$ and $1/\epsilon$
poles of the Passarino-Veltman functions, one can show that the divergent
contribution to the sum of all virtual graphs is again proportional to
the overall Born amplitude,
\beqn
\label{eq:box_tri_contrib}
{\cal M}_V &=& {\cal M}_B\frac{\alpha_s(\mu_R)}{4\pi} C_F
\(\frac{4\pi\mu_R^2}{Q^2}\)^\epsilon  \\
&\times& \Gamma(1+\epsilon)
\lq-\frac{2}{\epsilon^2}-\frac{3}{\epsilon}+\frac{\pi^2}{3}-7\rq 
+\tilde{\cal M}_V \;, \nonumber
\eeqn
where $\tilde{\cal M}_V$ is finite. Thus, the interference between the 
Born amplitude and the virtual correction amplitude, 
$2 \Re \lq {\cal M}_V{\cal M}_B^*\rq$, exactly cancels the divergent terms
in Eq.~(\ref{eq:I}).  The remaining integrals, involving the
%finite remainder $2 \Re \lq \tilde{\cal M}_V {\cal M}_B^*\rq$, 
finite remainder $2 \Re [ \tilde{\cal M}_V {\cal M}_B^*]$, 
are performed numerically in $D=4$ dimensions.

\begin{figure}[bht] 
\centerline{ 
\epsfig{figure=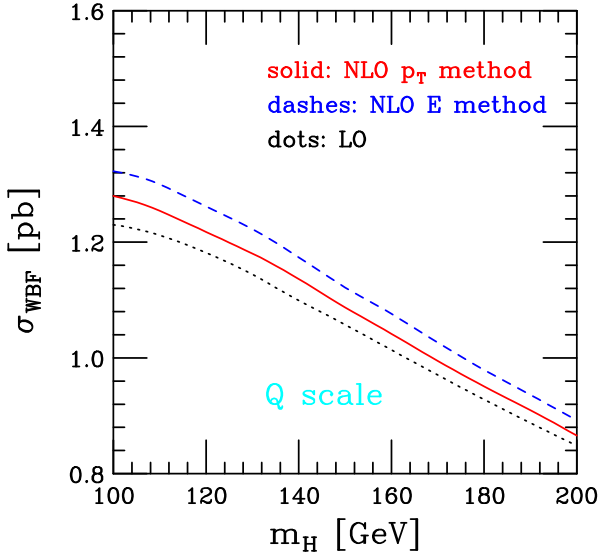,angle=90,width=0.5\textwidth,clip=} \ \  
} 
\vspace*{-0.25in}
\caption{ 
\label{fig:sigtotH}
Total Higgs boson production cross section as a function of the Higgs boson
mass,  
within the cuts described in the text at LO (dotted line) and at NLO. 
The NLO curves are shown for selection of the tagging jets 
as the two most energetic jets (E method) or as the two jets of highest
transverse momentum ($p_T$ method). 
}
\vspace*{-0.15in}
\end{figure} 

\section{Predictions for the LHC}
The calculations discussed above have been implemented in the form of a
NLO Monte Carlo program. This permits the determination of arbitrary infrared
and collinear safe distributions and cross sections within cuts. 
 In order to reconstruct jets from the final-state partons,
the $k_T$-algorithm~\cite{kToriginal} as described in Ref.~\cite{kTrunII} 
is used, with resolution parameter $D=0.8$. As a default, 
electroweak parameters are determined in the $G_\mu$ scheme, with
$m_Z=91.188$~GeV, $m_W=80.419$~GeV and the measured value of $G_F$
as our electroweak input, from which we obtain
$\alpha_{QED}=1/132.51$ and $\sin^2\theta_W=0.2223$, using LO electroweak
relations. The decay widths are then calculated as $\Gamma_W=2.099$~GeV and
$\Gamma_Z=2.510$~GeV.  We use the CTEQ6M parton distribution functions 
(PDFs)~\cite{cteq6} with $\alpha_s(m_Z)=0.118$ for all NLO results and
CTEQ6L1 parton distributions for all LO cross sections. For further
details, see Refs.~\cite{Figy:2003nv,Oleari:2003tc}.

In the following we consider $Hjj$, $Wjj$ and $Zjj$ cross sections
within generic cuts which are relevant for VBF studies at the LHC. We
ask for events with at least two hard jets 
which are required to have
\beq
\label{eq:cuts1}
p_{Tj} \geq 20~{\rm GeV} \, , \qquad\qquad |y_j| \leq 4.5 \, .
\eeq
Here $y_j$ denotes the rapidity of the (massive) jet momentum which is 
reconstructed as the four-vector sum of massless partons of 
pseudorapidity $|\eta|<5$. The two reconstructed jets of highest transverse 
momentum are called ``tagging jets'' and are identified with the final-state
quarks which are characteristic for VBF processes. 

We calculate cross sections for decays $Z\to \ell^+\ell^-$ and $W\to \ell\nu$
into a single generation of leptons. In order to ensure that the charged 
leptons are well observable, we impose the lepton cuts
\beq
\label{eq:cuts2}
p_{T\ell} \geq 20~{\rm GeV} \,,\quad |\eta_{\ell}| \leq 2.5  \,,\quad 
\triangle R_{j\ell} \geq 0.4 \, ,
\eeq
where $R_{j\ell}$ denotes the jet-lepton separation in the rapidity-azimuthal
angle plane. In addition, the charged leptons are required to fall between
the rapidities of the two tagging jets,
\beq
\label{eq:cuts3}
y_{j,min}  < \eta_\ell < y_{j,max} \, .
\eeq
For the case of Higgs boson production, the approximately massless Higgs boson
decay products $\tau^+\tau^-$, $\bar bb$ or $\gamma\gamma$ play the role
of the leptons and we impose the above lepton cuts on them, except that
$\triangle R_{j\ell} \geq 0.6$ is used in the Higgs boson case. Since the
branching ratio, $B$, for each of these final states depends on the respective
coupling to the Higgs boson, we show Higgs boson cross sections renormalized
to a branching ratio of 100\% for the selected final state, i.e.\ the
cross section within cuts is multiplied by an overall factor $1/B$.

Backgrounds to vector-boson fusion are significantly suppressed by requiring
a large rapidity separation of the two tagging jets, i.e.\ we require
\beq
\label{eq:cuts4}
\Delta y_{jj}=|y_{j_1}-y_{j_2}|>4\; .
\eeq

\begin{figure}[bht] 
\centerline{ 
\epsfig{figure=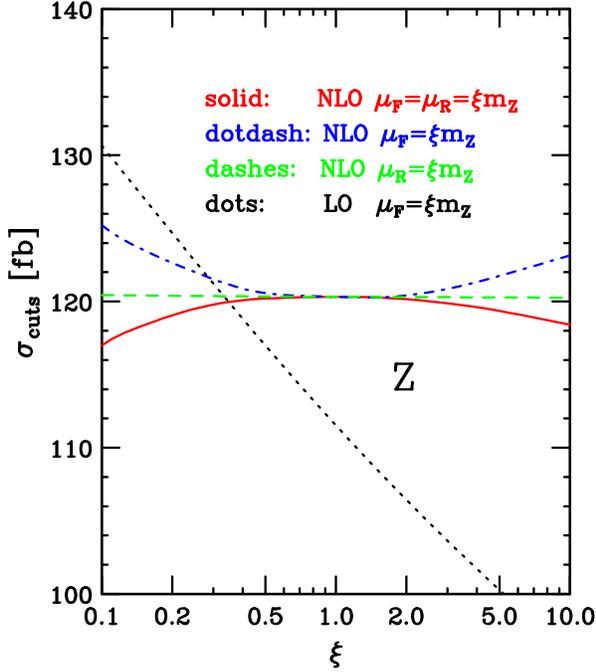,width=0.5\textwidth,clip=} \ \  
} 
\vspace*{-0.25in}
\caption{ 
\label{fig:scaleZ}
Scale dependence of the $Z\to \mu^+\mu^-$ cross section in VBF at the LHC. 
}
\vspace*{-0.15in}
\end{figure} 

The Higgs boson production cross section, within the above cuts, is shown in 
Fig.~\ref{fig:sigtotH} as a function of the Higgs boson mass. QCD corrections
increase the LO cross section (dotted line) slightly, by about 3 to 8\%,
depending on the Higgs boson mass. In addition to our default of identifying
the tagging jets as the jets of highest transverse momentum ($p_T$ method), 
Fig.~\ref{fig:sigtotH} also shows the NLO result which is obtained when 
defining the tagging jets as the two most energetic jets in a given
event ($E$ method, dashed line). From a consideration of tagging jet
$p_T$-distributions, the $p_T$ method appears to be somewhat more stable
against scale variations, however~\cite{Figy:2003nv}.

The typical scale dependence of the NLO VBF cross sections is
demonstrated in Fig.~\ref{fig:scaleZ}, where the $Zjj$ cross section
within the cuts of Eqs.~(\ref{eq:cuts1}--\ref{eq:cuts4}) is shown as a
function of the renormalization scale, $\mu_R=\xi m_Z$, holding the
factorization scale fixed at $\mu_F=m_Z$ (dashed curve), as a function
of $\mu_F=\xi m_Z$ for $\mu_R=m_Z$ (dot-dashed curve) or when varying
both. In all cases one finds scale variations at below the 2\% level
(when allowing variations between $\xi=0.5$ and $\xi=2$) which is minute
compared to the LO variation (dotted curve). This result holds for
Higgs, $W$ and $Z$ production as well as for the scale choice 
$\mu=\xi Q$, where $Q$ denotes the virtuality of the $t$-channel weak 
bosons in the various VBF processes.

\begin{figure}[!ht]
\vspace*{-0.03in}
\centerline{ 
\epsfig{figure=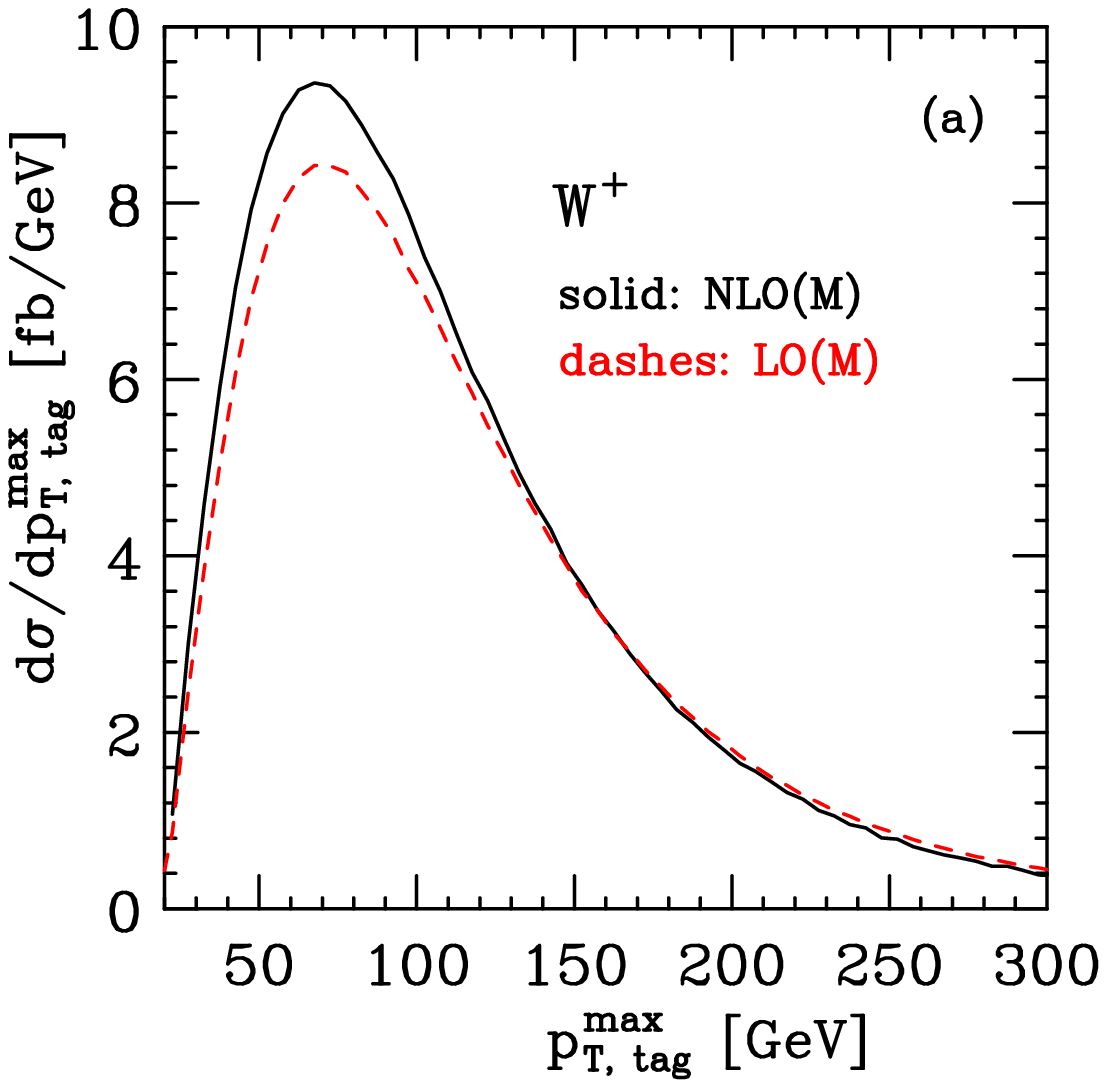,angle=90,width=0.47\textwidth,clip=} 
} 
\centerline{ 
\epsfig{figure=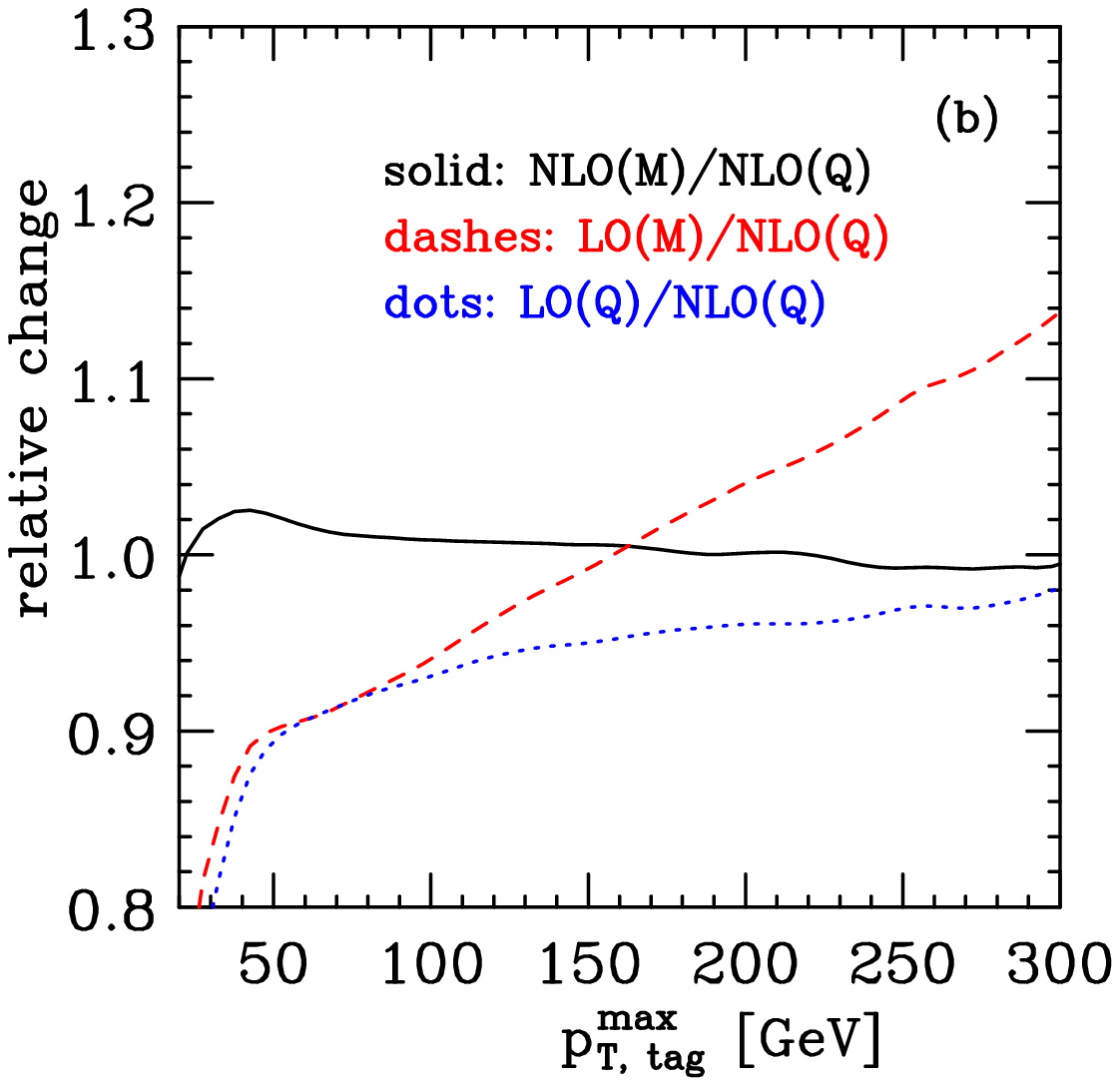,angle=90,width=0.47\textwidth,clip=} 
} 
\vspace*{-0.3in}
\caption{
\label{fig:ptmaxj}
Transverse-momentum distribution of the highest-$p_T$ 
jet in $W^+$ production at the LHC. In panel~(a) the differential
distribution is shown at LO and NLO for the scale choice
$\mu_F=\mu_R=m_W$ ($M$ scheme).
In panel~(b), we show the ratios of the NLO differential cross section in
the $M$ scheme (solid black line), of the LO one in the $M$ scheme (dashed
red line) and of the LO one in the $Q$ scheme (blue dotted line) 
to the NLO distribution for the scale choice $\mu_F=\mu_R=Q_i$.
}
\end{figure} 

The NLO QCD corrections to observable VBF cross sections are modest,
amounting to an increase of less than 10\% for $\xi=1$ in the examples 
discussed above. Similarly modest QCD corrections are found for most
distributions also. One example is shown in Fig.~\ref{fig:ptmaxj}, where
the tranverse momentum of the hardest tagging jet in $W^+$ events is
given. A comparison of the LO (dashed line) and NLO distribution (solid
line) in panel~(a) exhibits modest shape changes. They are
more clearly exposed in Fig.~\ref{fig:ptmaxj}(b) where the ratio of
distributions
\beq
R(p_T) = \frac{d\sigma^{(N)LO}(\mu)/dp_T}{d\sigma^{NLO}(\mu=Q)/dp_T}
\eeq
is shown, i.e.\ we compare to the default scale choice $\mu=Q$ at NLO.
Using
the $W$ mass as the renormalization and factorization scale instead
introduces changes below the 2\% level at NLO (solid line) which again
points to the stability of the NLO predictions. At LO
the differential cross section for this scale choice is suppressed by 10\%
and more at low $p_{T,\rm tag}^{\rm max}$, and is enhanced by up to 10\%
at high $p_T$, i.e.\ there is a substantial shape change from the  LO to
the NLO prediction. These changes are somewhat less pronounced when
choosing $\mu=Q$ for the LO calculation.

\section{Conclusions}

The one-loop QCD corrections to Higgs, $W$ and $Z$ production via VBF 
in hadronic collisions are now available in the form of a flexible NLO
parton-level Monte Carlo program. The NLO corrections are modest for
cross sections as well as differential distributions at the LHC, rarely
exceeding the 10\% level. In a variety of distributions one finds shape 
changes at the 10\% level. These changes need to be taken into account in
precision studies of VBF processes at the LHC.

\end{document}